\documentclass[fleqn,preprint,5p]{elsarticle}

\usepackage{amsmath,graphicx,amssymb}
\usepackage{epsf,psfig}

\journal{Mechanics Research Communications}

\begin{document}

\begin{frontmatter}

\title{A semi-numerical method for periodic orbits in a bisymmetrical potential}

\author{N. D. Caranicolas}
\author{E. E. Zotos\corref{}}

\address{Department of Physics, \\\
Section of Astrophysics, Astronomy and Mechanics, \\\
Aristotle University of Thessaloniki \\\
541 24, Thessaloniki, Greece}

\cortext[]{Corresponding author: \\\
\textit{E-mail address}: evzotos@astro.auth.gr (Euaggelos E. Zotos)}

\begin{abstract}

We use a semi-numerical method to find the position and period of periodic orbits in a bisymmetrical potential, made up of a two dimensional harmonic oscillator, with an additional term of a Plummer potential, in a number of resonant cases. The results are compared with the outcomes obtained by the numerical integration of the equations of motion and the agreement is good. This indicates that the semi-numerical method gives general and reliable results. Comparison with other methods of locating periodic orbits is also made.
\end{abstract}


\end{frontmatter}

\section{Introduction}

Dynamical systems made up of harmonic oscillators have been extensively used over the last five decades in order to describe motion  near an equilibrium point (see e.g H\'{e}non and Heiles, 1964; Giorgilli and Galgani, 1978; Saito and Ichimura, 1979; Caranicolas, 1993; 1994a; Caranicolas and Karanis, 1998; 1999; Arribas at al., 2006). In order to study these systems, scientists have used numerical (Karanis and Vozikis, 2008) or analytical methods (see Caranicolas and Barbanis, 1982; Deprit, 1991; Deprit and Elipe, 1991).

Astronomers frequently use potentials made up of harmonic oscillators, in order to study local motion in galaxies. Of particular interest are the bisymmetrical potentials, as these systems have been used in order to describe motion near the center of an elliptical galaxy. A large part of the above studies have been devoted to locate the position and find the period of the periodic orbits, as these orbits represent the backbone of the whole set of orbits.

In an earlier paper Caranicolas and Innanen (1992) (hereafter Paper I), we studied the periodic motion in  potentials of the form
\begin{flalign}
&V_I = \frac{1}{2}\left(\omega_1^2 x^2 + \omega_2^2 y^2 \right) + h.o.t,&
\end{flalign}
in the case where $\omega_1 = \omega_2$, that is when the frequencies of oscillations along the $x$ and $y$ axis, respectively, are equal. This is called a perturbed elliptic oscillator. In paper I, we used a combination of numerical and analytical calculations in order to find the position and the period of the periodic orbits for different $h.o.t$ (higher order terms), that is for different perturbation functions.

In this article we shall use the potential
\begin{flalign}
&V(x,y) = \frac{1}{2}\left(\omega_1^2 x^2 + \omega_2^2 y^2 \right) - \frac{\mu}{\sqrt{x^2 + y^2 + \alpha^2}},&
\end{flalign}
where $\mu$ and $\alpha$ are parameters. Potential (2) consists of a two dimensional harmonic oscillator with an additional Plummer sphere and can be used to describe motion in the central parts of a galaxy with a central bulge and a dense nucleus. The equations of motion for a test particle with a unit mass are
\begin{flalign}
&\ddot{x} = - \left[\omega_1^2 + \frac{\mu}{\left(x^2 + y^2 + \alpha^2 \right)^{3/2}}\right] x = - \omega_{1a}^2 x,& \nonumber \\
&\ddot{y} = - \left[\omega_2^2 + \frac{\mu}{\left(x^2 + y^2 + \alpha^2 \right)^{3/2}}\right] y = - \omega_{2a}^2 y,&
\end{flalign}
where the dot indicates derivative with respect to the time. The corresponding Hamiltonian is written as
\begin{flalign}
&H = \frac{1}{2}\left(p_x^2 + p_y^2 + \omega_1^2 x^2 + \omega_2^2 y^2 \right) - \frac{\mu}{\sqrt{x^2 + y^2 + \alpha^2}} = h,&
\end{flalign}
where $p_x$ and $p_y$ are the momenta per unit mass conjugate to $x$ and $y$, while $h > 0$ is the numerical value of the Hamiltonian, which is conserved. Our aim is to find the position of periodic orbits and the corresponding periods in the resonant cases, using a semi-numerical procedure and also to compare the results with the outcomes given by the numerical integration of the equations of motion. We shall study the resonant cases $\omega_1$:$\omega_2$ = 1:1, $\omega_1$:$\omega_2$ = 2:1, $\omega_1$:$\omega_2$ = 2:3 and $\omega_1$:$\omega_2$ = 4:3. In order to keep thing simple, we use the values: $\mu = 0.001$ and $\alpha = 0.25$, while the value of the energy $h$, will be treated as a parameter. For the numerical integration of the equations of motion, a Bulirsh-St\"{o}er numerical integration routine in double precision is used. The accuracy of the calculations is checked by the constancy of the energy integral (4), which is conserved up to the twelfth significant figure.

The paper is organized as follows: In Section 2 we find the position and the periods of the periodic orbits starting perpendicularly from the $x$ axis. In Section 3 we study periodic orbits going through the origin. A discussion and the conclusions of this work are presented in Section 4.

\section{Periodic orbits starting perpendicularly from the $x$ axis}

We shall start for case when $\omega_1 = \omega_2 = \omega$, that is for the 1:1 resonant case. In this case the potential (2) is axially symmetric and in polar coordinates $\left(r,\phi\right)$ takes the form
\begin{flalign}
&V(r) = \frac{1}{2}\omega^2 r^2 - \frac{\mu}{\sqrt{r^2 + \alpha^2}}.&
\end{flalign}
As the potential is axially symmetric the test particle's angular momentum $L$ is an integral of motion. In this case the system has a circular periodic orbit of radius $r_c$. Elementary calculations show that the angular momentum of this circular orbit is given by the equation
\begin{flalign}
&L_c^2 = r_c^3 \left[\frac{dV}{dr}\right]_{r_c},&
\end{flalign}
while the corresponding value of the energy is
\begin{flalign}
&h_c = \frac{1}{2}\frac{L_c^2}{r_c^2} + V(r).&
\end{flalign}

The period of the circular periodic orbit can be found, if we observe that as the orbit is a circle with a radius $r_c$ the two frequencies $\omega_{1a}$ and $\omega_{2a}$ in equations (3) are always equal with a value
\begin{flalign}
&\omega_{1a} = \omega_{2a} = \omega_c = \left[\omega^2 + \frac{\mu}{\left(r_c^2 + \alpha^2\right)^{3/2}}\right]^{1/2}.&
\end{flalign}
Thus the period of the circular orbit is given by
\begin{flalign}
&T_c = \frac{2 \pi}{\left[\omega^2 + \frac{\mu}{\left(r_c^2 + \alpha^2 \right)^{3/2}}\right]^{1/2}}.&
\end{flalign}

We now proceed to find the position of the periodic orbits in the case when $\omega_1 = 2 \omega_2$, that is in the resonance case 2:1. Let $x_{s0}$ be the starting position of the periodic orbit from the $x$ axis. Remember that $y_0 = p_{x0} = 0$, while the value of $p_{y0}$ is always found from the energy integral (4). Our numerical calculations show that the above value of $p_{y0}$ is in a good  agreement with the value
\begin{flalign}
&p_{y0} = \frac{x_{s0}}{\omega_2}.&
\end{flalign}
If we set $x_{s0} = x, y = p_x = 0, p_y = x_{s0} / \omega_2$, in the harmonic part of the Hamiltonian (4), that is when $\mu = 0$ and solve the corresponding equation for $x_s$, we find
\begin{flalign}
&x_{s0} = \left[\frac{2h\omega_2^2}{1+\omega_1^2 \omega_2^2}\right]^{1/2},&
\end{flalign}
which gives a first value for the starting point of the periodic orbit. Numerical experiments indicate that a good approximation for the frequency of the 2:1 periodic orbits is
\begin{flalign}
&w_{1s} = \left[\omega_1^2 + \frac{\mu}{2\left(x_{s0}^2 + \alpha^2\right)^{3/2}}\right]^{1/2}.&
\end{flalign}
Setting this value of the frequency in equation (10) we find
\begin{flalign}
&p_{ys} = \frac{x_{s}}{w_{1s}},&
\end{flalign}
where $x_s$ is the new value of the starting point of the 2:1 periodic orbit. We now set $x_s = x, y = p_x = 0, p_{ys} = x_s / w_{2s}, w_{2s} = w_{1s} / 2$ in  Hamiltonian (4) and get
\begin{flalign}
&\frac{1}{2}\left(w_{1s}^2 + \frac{1}{w_{2s}^2}\right) x_s^2 - \frac{\mu}{\sqrt{x_s^2 + \alpha^2}} = h.&
\end{flalign}
If we solve equation (14) for $x_s$ we obtain the starting point of the 2:1 periodic orbit. The period of the 2:1 periodic orbit is given by
\begin{flalign}
&T_s = \frac{4 \pi}{w_{1s}}.&
\end{flalign}

Table 1 gives the positions and periods of the 2:1 periodic orbits for several values of the energy $h$. Subscript $n$ indicates values found by numerical integration, while subscript $s$ indicates values found using equations (14) and (15). One can see that the differences in the position of periodic orbits vary from 0 to about $9\%$, while in all cases the differences in the period are less than $1\%$. The values of the parameters are: $\omega_1 = 0.8, \omega_2 = 0.4, \mu = 0.001, \alpha = 0.25$. Figure 1 shows a 2:1 periodic orbit, for the above values of parameters when $h = 0.05$. The initial conditions are: $x_0 = x_n = 0.1295, y_0 = p_{x0} = 0$, while the value of $p_{y0}$ is found from the energy integral (4). The outermost curve is the curve of zero velocity.
\begin{table}
\centering
\caption{Positions and periods of the 2:1 resonant periodic orbits intersecting the $x$ axis perpendicularly. The values of the parameters are $\omega_1 = 0.8$ and $\omega_2 = 0.4$. Subscript $s$ indicates values derived by semi-numerical methods, while subscript $n$ shows results obtained by numerical integration.}
\begin{tabular}{|c||c|c|c|c|}
\hline
$h$   & $x_s$  & $x_n$  & $T_s$   & $T_n$ \\
\hline \hline
0.049 & 0.1254 & 0.1254 & 15.4270 & 15.5148 \\
\hline
0.050 & 0.1265 & 0.1295 & 15.4285 & 15.5183 \\
\hline
0.051 & 0.1276 & 0.1335 & 15.4300 & 15.5217 \\
\hline
0.052 & 0.1288 & 0.1373 & 15.4316 & 15.5249 \\
\hline
0.053 & 0.1300 & 0.1410 & 15.4331 & 15.5281 \\
\hline
0.054 & 0.1311 & 0.1446 & 15.4345 & 15.5312 \\
\hline
\end{tabular}
\end{table}
\begin{figure}[!tH]
\centering
\resizebox{\hsize}{!}{\rotatebox{0}{\includegraphics*{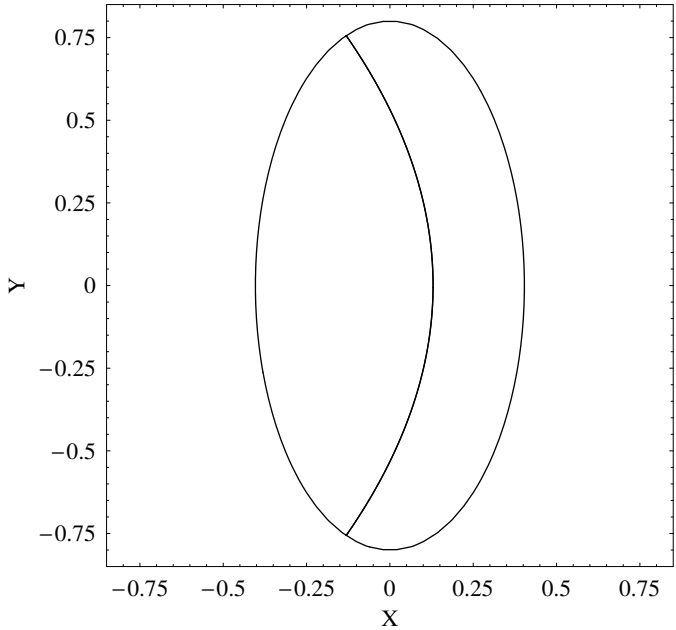}}}
\caption{A 2:1 resonant periodic orbit intersecting the $x$ axis perpendicularly. The values of parameters are as in Table 1. The value of energy is $h = 0.05$. Initial conditions are: $x_0 = x_n = 0.1295, y_0 = p_{x0} = 0$, while the value of $p_{y0}$ is found from the energy integral.}
\end{figure}

Next we come to study the 2:3 resonance, that is when $\omega_1$ : $\omega_2$ = 2:3. Here the numerical calculations suggest that a good approximation for the starting point of the periodic orbit is given by the formula
\begin{flalign}
&x_s = \frac{1}{\omega_1}\sqrt{\frac{\omega_2}{\omega_1}h},&
\end{flalign}
while the value of the corresponding period is
\begin{flalign}
&T_s = \frac{4\pi}{w_{1s}},&
\end{flalign}
where
\begin{flalign}
&w_{1s} = \left[\omega_1^2 + \frac{\mu}{\left(x_s^2 + \alpha^2\right)^{3/2}}\right]^{1/2}.&
\end{flalign}

Table 2 is similar to Table 1, but for the 2:3 periodic orbits. We see that the agreement between the results given by numerical integration and the outcomes from the semi-numerical formulas (16) and (17) is very good. The values of the parameters are: $\omega_1 = 0.4, \omega_2 = 0.6, \mu = 0.001, \alpha = 0.25$. Figure 2 shows a 2:3 periodic orbit, for the above values of parameters when $h = 0.08$. Initial conditions are: $x_0 = x_n = 0.8562, y_0 = p_{x0} =0$, while the value of $p_{y0}$ is found from the energy integral (4).
\begin{table}
\centering
\caption{Similar to Table 1 but for the 2:3 resonant periodic orbits. The values of the parameters are: $\omega_1 = 0.4, \omega_2 = 0.6$.}
\begin{tabular}{|c||c|c|c|c|}
\hline
$h$   & $x_s$  & $x_n$  & $T_s$   & $T_n$ \\
\hline \hline
0.065 & 0.7806 & 0.7832 & 31.2391 & 31.1167 \\
\hline
0.070 & 0.8100 & 0.8096 & 31.2560 & 31.1379 \\
\hline
0.075 & 0.8385 & 0.8334 & 31.2704 & 31.1578 \\
\hline
0.080 & 0.8660 & 0.8562 & 31.2827 & 31.1755 \\
\hline
0.085 & 0.8926 & 0.8789 & 31.2934 & 31.1911 \\
\hline
0.090 & 0.9185 & 0.9002 & 31.3027 & 31.2053 \\
\hline
\end{tabular}
\end{table}
\begin{figure}[!tH]
\centering
\resizebox{\hsize}{!}{\rotatebox{0}{\includegraphics*{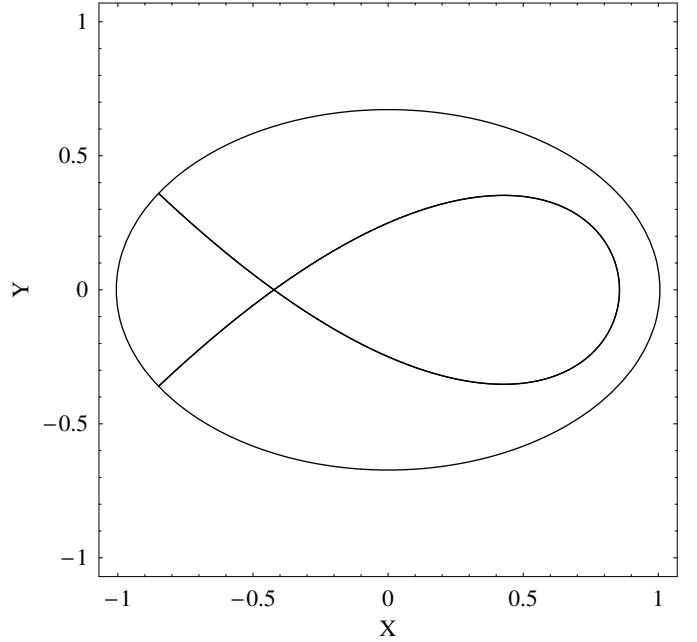}}}
\caption{A 2:3 resonant periodic orbit intersecting the $x$ axis perpendicularly. The values of parameters are as in Table 2. The value of energy is $h = 0.08$. Initial conditions are: $x_0 = x_n = 0.8562, y_0 = p_{x0} = 0$, while the value of $p_{y0}$ is found from the energy integral.}
\end{figure}

Finally we shall study the case when $\omega_1$ : $\omega_2$ = 4:3. In this case our numerical experiments show that the value of the starting point of the periodic orbit is well approximated if we use the formula (16). The value of the corresponding period is now given by the formula
\begin{flalign}
&T_s = \frac{8\pi}{w_{1s}},&
\end{flalign}
where
\begin{flalign}
&w_{1s} = \left[\omega_1^2 + \frac{\mu}{2\left(x_s^2 + \alpha^2\right)^{3/2}}\right]^{1/2}.&
\end{flalign}

In Table 3 we make a comparison between the results given by the numerical integration and those using formulas (16) and (19) for several values of the energy $h$. We see that the agreement is again very good. In all numerical calculations we use the values: $\omega_1 = 0.8, \omega_2 = 0.6, \mu = 0.001, \alpha = 0.25$. Figure 3 shows a 4:3 periodic orbit, for the above values of parameters, when $h=0.40$. Initial conditions are: $x_0 = x_n = 0.6847, y_0 = p_{x0} = 0$, while the value of $p_{y0}$ is found from the energy integral (4).
\begin{table}
\centering
\caption{Similar to Table 1 but for the 4:3 resonant periodic orbits. The values of the parameters are: $\omega_1 = 0.8, \omega_2 = 0.6$.}
\begin{tabular}{|c||c|c|c|c|}
\hline
$h$  & $x_s$  & $x_n$  & $T_s$   & $T_n$ \\
\hline \hline
0.35 & 0.6404 & 0.6401 & 31.3782 & 31.3684 \\
\hline
0.40 & 0.6846 & 0.6847 & 31.3843 & 31.3762 \\
\hline
0.45 & 0.7261 & 0.7266 & 31.3889 & 31.3820 \\
\hline
0.50 & 0.7654 & 0.7656 & 31.3924 & 31.3865 \\
\hline
0.55 & 0.8028 & 0.8027 & 31.3953 & 31.3901 \\
\hline
0.60 & 0.8385 & 0.8389 & 31.3976 & 31.3931 \\
\hline
\end{tabular}
\end{table}
\begin{figure}[!tH]
\centering
\resizebox{\hsize}{!}{\rotatebox{0}{\includegraphics*{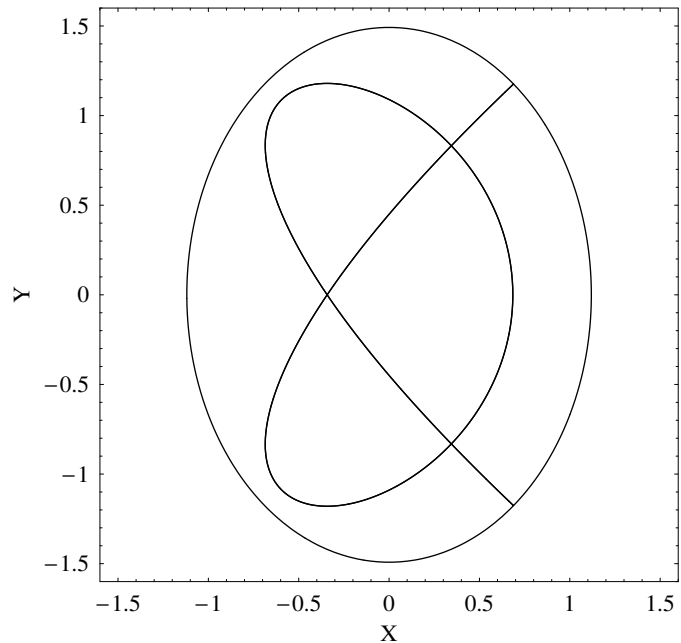}}}
\caption{A 4:3 resonant periodic orbit intersecting the $x$ axis perpendicularly. The values of parameters are as in Table 3. The value of energy is $h = 0.40$. Initial conditions are: $x_0 = x_n = 0.6847, y_0 = p_{x0} = 0$, while the value of $p_{y0}$ is found from the energy integral.}
\end{figure}

From the study of the periodic orbits starting perpendicularly from the $x$ axis in the above four resonance cases we can say the following. In the 1:1 resonance case there are analytical results for the position and period of the periodic orbits. On the other hand, in the other three resonance cases the semi-numerically derived formulas give interesting results. The outcomes from these formulas are in very good agreement with the results from numerical integration in the cases of the 2:3 and 4:3 resonant cases, while in the case of the 2:1 resonance the observed deviations are smaller than $10\%$.

\section{Periodic orbits going through the origin}

In this Section we shall study the periodic orbits going through the origin. We start again from the he 1:1 interesting resonant case, when $\omega_1 = \omega_2 = \omega$. For the motion along the lines
\begin{flalign}
&x = \pm \lambda y,&
\end{flalign}
the equations of motion become identical, while the frequencies of the oscillations are always equal. Thus the straight lines (21) are exact 1:1 resonant periodic orbits going though the origin. The value of the corresponding period can be found if we use the formula
\begin{flalign}
&T_s = \frac{2\pi}{w_s}.&
\end{flalign}

The frequency $w_s$ is given by the equation
\begin{flalign}
&w_s = \left[\omega^2 + \frac{0.8 \mu}{\left(r_a^2/2 + \alpha^2\right)^{3/2}}\right]^{1/2},&
\end{flalign}
where $r_a$ is the radius of the circle $V(r) = h$. Therefore all iso-energetic orbits (21) have the same period. In Table 4 we make a comparison between the period $T_s$ and the period $T_n$ found using numerical integration, for the 1:1 straight line periodic orbits, for six values of the energy $h$. As one can see there is a good agreement between $T_s$ and $T_n$.
\begin{table}
\centering
\caption{Periods of the 1:1 periodic orbits going through the origin. The values of the parameters are: $\omega_1 = \omega_2 = 0.4$.}
\begin{tabular}{|c||c|c|}
\hline
$h$  & $T_s$   & $T_n$ \\
\hline \hline
0.02 & 15.2813 & 15.2839 \\
\hline
0.04 & 15.4961 & 15.4866 \\
\hline
0.06 & 15.5770 & 15.5644 \\
\hline
0.08 & 15.6171 & 15.6043 \\
\hline
0.10 & 15.6403 & 15.6280 \\
\hline
0.12 & 15.6551 & 15.6436 \\
\hline
\end{tabular}
\end{table}

In the case of the 2:1 resonance we did not find periodic orbits going through the origin. In fact, we searched up to values of the energy five times as large as those of Table 2, without finding any 2:1 periodic orbits going through the origin.

Thus we come to find the periodic orbits going through the origin in the 2:3 resonance. The numerical calculations support the idea  that the orbit starts from the origin with
\begin{flalign}
&p_{xs} = \sqrt{\frac{\omega_2}{\omega_1}\left(h + \frac{\mu}{\alpha}\right)},&
\end{flalign}
while the value of $p_{ys}$ is found from the energy integral (4). At the origin the system behaves as separable. The energy $E_{xs}$ along the $x$ axis is
\begin{flalign}
&E_{xs} = \frac{1}{2}p_{xs}^2.&
\end{flalign}
Thus
\begin{flalign}
&E_{ys} = h - E_{xs} = h - \frac{1}{2}p_{xs}^2.&
\end{flalign}

We also observe that these orbits intersect the $y$ axis perpendicularly. Let $y_s$ be the point of intersection. Numerical experiments show that the point of intersection can be found with sufficient accuracy using the equation
\begin{flalign}
&y_s = \left(\frac{2h - p_{xs}^2}{\omega_2^2}\right)^{1/2}.&
\end{flalign}

The period of the 2:3 periodic orbits going through the origin is well approximated if we use the formula
\begin{flalign}
&T_s = \frac{4\pi}{\left[\omega_1^2 + \frac{\mu}{10\left(y_s^2 + \alpha^2\right)^{3/2}}\right]^{1/2}}.&
\end{flalign}

Table 5 gives the values of the starting positions $p_{xs}$ and $p_{xn}$ of the 2:3 periodic orbits going through the origin. In the same Table, we give the values of $y_s$ and $y_n$ and the corresponding periods $T_s$ and $T_n$ for six values of the energy $h$. The agreement is good. The values of the parameters and the energies are as in Table 3. Figure 4 shows a 2:3 periodic orbit going through the origin for the above values of parameters when $h=0.09$. Initial conditions are: $x_0 = y_0 = 0, p_{xn} = 0.3768$, while the value of $p_{y0}$ is found from the energy integral (4).
\begin{table}
\centering
\setlength{\tabcolsep}{4.0pt}
\caption{Positions and periods of the 2:3 resonant periodic orbits going through the origin. The values of the parameters are: $\omega_1 = 0.4, \omega_2 = 0.6$. Subscript $s$ indicates values derived by semi-numerical methods, while subscript $n$ shows results obtained by numerical integration.}
\begin{tabular}{|c||c|c|c|c|c|c|}
\hline
$h$ & $p_{xs}$ & $p_{xn}$ & $y_s$ & $y_n$ & $T_s$   & $T_n$ \\
\hline \hline
0.065 & 0.3217 & 0.3319 & 0.2713 & 0.2652 & 31.2222 & 31.1287 \\
\hline
0.070 & 0.3331 & 0.3421 & 0.2838 & 0.2803 & 31.2360 & 31.1513 \\
\hline
0.075 & 0.3442 & 0.3520 & 0.2958 & 0.2947 & 31.2483 & 31.1711 \\
\hline
0.080 & 0.3549 & 0.3618 & 0.3073 & 0.3079 & 31.2592 & 31.1885 \\
\hline
0.085 & 0.3653 & 0.3672 & 0.3184 & 0.3331 & 31.2690 & 31.2016 \\
\hline
0.090 & 0.3755 & 0.3768 & 0.3291 & 0.3444 & 31.2778 & 31.2157 \\
\hline
\end{tabular}
\end{table}
\begin{figure}[!tH]
\centering
\resizebox{\hsize}{!}{\rotatebox{0}{\includegraphics*{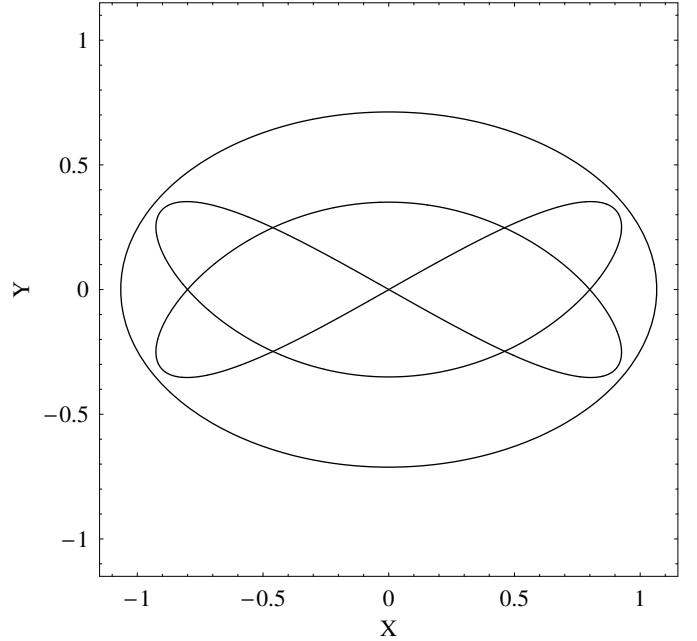}}}
\caption{A 2:3 periodic orbit going through the origin, when $h = 0.09$. The values of parameters are as in Table 2. Initial conditions are: $x_0 = y_0 = 0, p_{x0} = p_{xn} = 0.3768$, while the value of $p_{y0}$ is found from the energy integral.}
\end{figure}

Our last but not least case, is to give semi-numerical results for the position of the 4:3 periodic orbits going through the origin. It is found that the starting position of these orbits is given with good agreement if we use the following value for the $p_x$
\begin{flalign}
&p_{xs} = \sqrt{\frac{\omega_2}{\omega_1}h},&
\end{flalign}
while the value of $p_{ys}$ is found from the energy integral (4). Working as in the case of the 2:3 resonance we find that the $y_s$ can be obtained by the same formula (27), while a good approximation for the period is obtained if we use the equation
\begin{flalign}
&T_s = \frac{8\pi}{\left[\omega_1^2 + \frac{\mu}{2\left(y_s^2 + \alpha^2\right)^{3/2}}\right]^{1/2}}.&
\end{flalign}

Table 6 is similar to Table 5 for the 4:3 periodic orbits going through the origin. The values of the parameters and the energies are as in Table 3. Figure 5 shows a 4:3 periodic orbit going through the origin for the above values of parameters when $h=0.50$. Initial conditions are: $x_0 = y_0 = 0, p_{xn} = 0.6021$, while the value of $p_{y0}$ is found from the energy integral (4).
\begin{table}
\centering
\setlength{\tabcolsep}{4.0pt}
\caption{Similar to Table 5 for the 4:3 resonant periodic orbits. The values of the parameters are: $\omega_1 = 0.8, \omega_2 = 0.6$.}
\begin{tabular}{|c||c|c|c|c|c|c|}
\hline
$h$ & $p_{xs}$ & $p_{xn}$ & $y_s$ & $y_n$ & $T_s$   & $T_n$ \\
\hline \hline
0.35 & 0.5123 & 0.4975 & 1.1024 & 1.1232 & 31.4074 & 31.3698 \\
\hline
0.40 & 0.5477 & 0.5373 & 1.1785 & 1.1934 & 31.4089 & 31.3775 \\
\hline
0.45 & 0.5809 & 0.5725 & 1.2500 & 1.2624 & 31.4100 & 31.3830 \\
\hline
0.50 & 0.6123 & 0.6021 & 1.3176 & 1.3317 & 31.4108 & 31.3873 \\
\hline
0.55 & 0.6422 & 0.6326 & 1.3819 & 1.3957 & 31.4115 & 31.3907 \\
\hline
0.60 & 0.6708 & 0.6624 & 1.4433 & 1.4553 & 31.4120 & 31.3935 \\
\hline
\end{tabular}
\end{table}
\begin{figure}[!tH]
\centering
\resizebox{\hsize}{!}{\rotatebox{0}{\includegraphics*{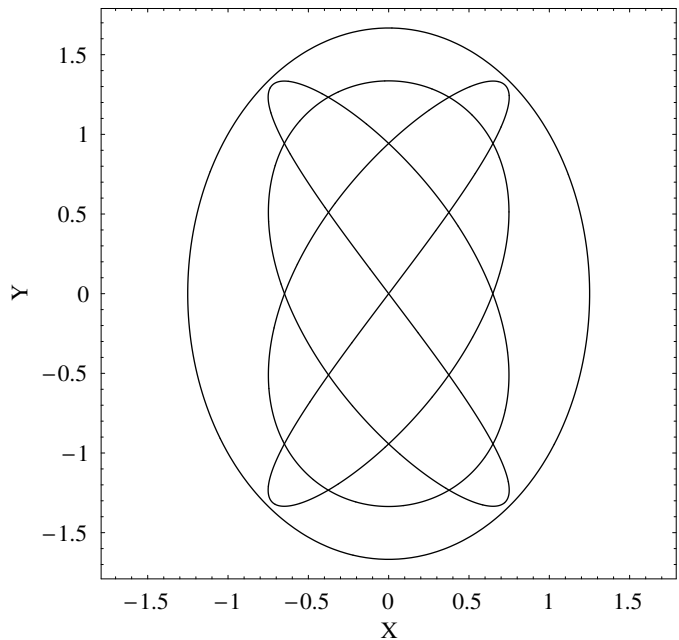}}}
\caption{A 4:3 periodic orbit going through the origin, when $h = 0.50$. The values of parameters are as in Table 3. Initial conditions are: $x_0 = y_0 = 0, p_{x0} = p_{xn} = 0.6021$, while the value of $p_{y0}$ is found from the energy integral.}
\end{figure}

In this Section we have studied the orbits going through the origin. Here we presented the resonant cases 1:1, 2:3 and 4:3. In the resonant case 2:1 we did not find resonant periodic orbits going through the origin. Our study shows that the semi-numerical formulas can give good a approximation for the position and periods of the resonant periodic orbits.

\section{Discussion}

In this paper we have studied the periodic orbits in a dynamical system composed of a two dimensional harmonic oscillator and a Plummer potential in the resonant cases $\omega_1$ : $\omega_2$ = $n/m$ = 1:1, 2:1, 2:3 and 4:3. We chose the above resonances because all of them are important resonances. On the other hand, it is not possible to study all resonances in a given potential.

In this potential we presented semi-numerical formulas for the position and the period of the resonant periodic orbits intersecting the $x$ axis perpendicularly and going through the origin. In semi-numerical methods we combine empirical or theoretical results or both, with results found numerically, in order to find relations or formulas giving outcomes that are close to those obtained by pure numerical methods, that is by numerical integration. At this point, it would be useful to refer to how the numerical results indicate or suggest the formulas giving the position and period of the periodic orbits. In other words, to explain how the formulas are obtained. This is done on a complete empirical basis, using simple expressions which, of course, contain the parameters entering the Hamiltonian (4) and give outcomes that are close to those given by the numerical integration. If the agreement is not satisfactory, we look for a new formula and so on, until we have achieved a good result. In some cases, we combine empirical and theoretical results in order to obtain a better accuracy (see equations (11)-(14)).

We have presented semi-numerical results in previous papers (see Paper I; Caranicolas, 1994a,b) not only for polynomial potentials representing local galactic motion (i.e. near an equilibrium point) but also for global galactic motion (Caranicolas and Innanen, 1991). Semi-numerical methods were also used in Celestial Mechanics (see Henrard and Caranicolas, 1990). Therefore these methods are a sharp tool in the study of dynamical systems.

It is well known that for the study of dynamical systems, scientists have used numerical or analytical methods (Deprit, 1991; Deprit and Elipe, 1991; Elipe, 2000; Elipe and Deprit, 1999), or the integrals of motion (Ferrer et al., 1997; Ferrer et al., 1998; Ferrer et al., 1998) or maps (Caranicolas, 1990). But analytical methods are not easy to apply in non-polynomial potentials (i.e. global galactic potentials such as mass models). Moreover maps or formal integrals of motion are not easy to find for  potentials of the form (1). Therefore we decided to apply the semi-numerical method in the potential (1). The good agreement between the results of the semi-numerical method and the outcomes given by the numerical integration of the equations of motion justifies our choice.

\section*{Acknowledgment}

\textit{We would like to express our thanks to the two anonymous referees for their very useful suggestions and comments, which improved the quality of the present paper.}


\section*{References}


\begin{thebibliography}{}


\bibitem{} Arribas, M., Elipe, A., Floria, L., Riaguas, A., 2006. Chaos, Solitons \& Fractals 27, 1220.

\bibitem{} Caranicolas, N.D., 1990. Celestial Mechanics 47, 87.

\bibitem{} Caranicolas, N.D., Karanis, G. I., 1998. A\&SS 259, 45.

\bibitem{} Caranicolas, N.D., Karanis, G. I., 1999. A\&A 342, 349.

\bibitem{} Caranicolas, N.D., Barbanis, B., 1982. A\&A 114, 360.

\bibitem{} Caranicolas, N.D., Innanen, K.A, 1991. AJ 102(4), 1343.

\bibitem{} Caranicolas, N.D., Innanen, K.A, 1992. AJ 103(4), 1308 (Paper I).

\bibitem{} Caranicolas, N.D., 1993. A\&A 267, 388.

\bibitem{} Caranicolas, N.D., 1994a. A\&A 282, 34.

\bibitem{} Caranicolas, N.D., 1994b. A\&A 287, 752.

\bibitem{} Deprit, A., 1991. Celestial Mechanics and Dynamical Astronomy 51, 202.

\bibitem{} Deprit, A., Elipe, A., 1991. Celestial Mechanics and Dynamical Astronomy 51, 227.

\bibitem{} Elipe, A., 2000. Physical Review E 61, 6477.

\bibitem{} Elipe, A., Deprit, A., 1999. Mechanics Research Communications 26, 635.

\bibitem{} Ferrer, S., Lara, M., et al., 1997. Physics Letters A 228, 255.

\bibitem{} Ferrer, S., Lara, M., et al., 1998a. International Journal of Bifurcations and Chaos 6, 1199.

\bibitem{} Ferrer, S., Lara, M., et al., 1998b. International Journal of Bifurcations and Chaos 8, 1215.

\bibitem{} Giorgilli, A., Galgani, L., 1978. Celestial Mechanics 17, 267.

\bibitem{} H\'{e}non, M., Heiles, C., 1964. AJ 69, 73.

\bibitem{} Henrard, J., Caranicolas, N.D., 1990. Celestial Mechanics 47, 99.

\bibitem{} Karanis, G.I., Vozikis, Ch.L., 2008. AN 323, 3.

\bibitem{} Saito, N., Ichimura, A., 1979. In: Casati, G., Ford, L. (Eds.), \textit{Stochastic Behaviour in Classical And Quantum Hamiltonian Systems}, Springer, Berlin.

\end{thebibliography}
\end{document}